# Radial spin texture in elemental tellurium with chiral crystal structure

M. Sakano[1,2], M. Hirayama[3,4,5], T. Takahashi[6], S. Akebi[1], M. Nakayama[1], K. Kuroda[1], K. Taguchi[7], T. Yoshikawa[7], K. Miyamoto[8], T. Okuda[8], K. Ono[9], H. Kumigashira[9,10], T. Ideue[2], Y. Iwasa[2,5], N. Mitsuishi[2], K. Ishizaka[2,5], S. Shin[1,11], T. Miyake[12], S. Murakami[3,4], T. Sasagawa[6], and Takeshi Kondo[1,11]

[1]Institute for Solid State Physics (ISSP), The University of Tokyo, Kashiwa, 277-8581, Japan
[2]Quantum-Phase Electronics Center (QPEC) and Department of Applied Physics, The University of Tokyo, Bunkyo-ku, Tokyo, 113-8656, Japan
[3]Department of Physics, Tokyo Institute of Technology, Meguro-ku, Tokyo, 152-8551, Japan
[4]Tokodai Institute for Element Strategy (TIES), Tokyo Institute of Technology, Meguro-ku, Tokyo, 152-8551, Japan
[5]RIKEN Center for Emergent Matter Science (CEMS), Wako, Saitama, 351-0198, Japan
[6]Materials and Structures Laboratory (MSL), Tokyo Institute of Technology, Yokohama, Kanagawa, 226-8503, Japan.
[7]Graduate School of Science, Hiroshima University, Higashi-Hiroshima, Hiroshima, 739-8526, Japan
[8]Hiroshima Synchrotron Radiation Center (HiSOR), Hiroshima University, Higashi-Hiroshima, Hiroshima, 739-0046, Japan
[9]Institute of Materials Structure Science, High Energy Accelerator Research Organization (KEK), Tsukuba, Ibaraki 305-0801, Japan
[10]Institute of Multidisciplinary Research for Advanced Materials (IMRAM), Tohoku University, Sendai 980-8577, Japan
[11]AIST-UTokyo Advanced Operando-Measurement Technology Open Innovation Laboratory (OPERANDO-OIL), Kashiwa, Chiba 277-8581, Japan
[12] Research Center for Computational Design of Advanced Functional Materials (CD-FMat), AIST, Tsukuba, Ibaraki, 305-8568, Japan

The chiral crystal is characterized by a lack of mirror symmetry and an inversion center, resulting in the inequivalent right- and left-handed structures. In the noncentrosymmetric crystal structure, the spin and momentum of electrons are locked in the reciprocal space with the help of the spin-orbit interaction. To reveal the spin textures of chiral crystals, here we investigate the spin and electronic structure in *p*-type semiconductor elemental tellurium with a chiral crystal structure by using spin- and angle-resolved photoemission spectroscopy. Our data demonstrate that the highest valence band crossing the Fermi level has a spin component parallel to the electron momentum around the BZ corners. Significantly, we have also confirmed that the spin polarization is reversed in the crystal with the opposite chirality. The results indicate that the spin textures of the right- and left-handed chiral crystals are hedgehog-like, leading to unconventional magnetoelectric effects and nonreciprocal phenomena.

Knowledge of crystal lattice symmetries helps us in understanding the physical phenomena and properties of materials. In a chiral crystal, it has long been known that the electricity and magnetism are coupled; in 1811[1], Arago discovered the natural optical activity in alpha-quartz with the representative chiral crystal. The symmetry of the chiral crystal structure is defined by the lack of mirror symmetry in itself, giving rise to the existence of two kinds of crystal structures connected to each other by the mirror operation or inversion operation. Such two inequivalent crystals, so-called right- and left-handed crystals, show different biological, chemical, and physical responses in nature. For example, D-glucose can be used as a source of energy in the human body, whereas L-glucose, which is the enantiomer of D-glucose, cannot[2]. The physical properties of chiral crystals are also intriguing in that the spin degeneracy should be lifted in momentum space (**k**-space), even in nonmagnetic materials, by the combination of the spin-orbit interaction (SOI) and the breaking of inversion symmetry. Investigations of the spin-split bands in the noncentrosymmetric crystals are originated in the works of Dresselhaus[3] and Rashba[4,5], who evaluated the effect of the SOI on the band dispersions in zinc blende and wurtzite semiconductors characterized by regular tetrahedron and polar crystal structures. The spin-momentum locked electronic structure allows one to exploit the spin degree of freedom of electrons even without the magnetism or external magnetic field. Therefore, the investigation of SOI in noncentrosymmetric materials has attracted particular attention in the rapidly growing field of spintronics.

The recent progress in the techniques of the spin- and angle-resolved photoemission spectroscopy (SARPES)[6-8] made it possible to three-dimensionally detect the in-plane and out-of-plane spin polarization of electrons, and the spin texture of band structure has been determined for various materials with a polar structure[9] or a three-fold rotational symmetry[10]. However, the direct observation of spin texture in the band structure of chiral crystal has been still lacking and long been desired in material science. The chiral materials must be distinguished from noncentrosymmetric materials investigated so far in that the inversion operation cannot transform the chiral crystal onto itself. In other words, while the sign of spin signal due to SOI can be reversed only by the rotation operation in a noncentrosymmetric non-chiral material with the mirror symmetry, it is not the case for the chiral material, where the interchange between right- and left-handed crystals is necessary to reverse its spin structure. This symmetry requirement distinctive in the chiral crystal should yield peculiar spin and electronic structures directly tied to spin-related physical properties.

Several calculational studies on the band structure of nonmagnetic chiral materials have predicted that the spin components parallel to the momentum of electrons should appear all-around the highly symmetrical **k**-points[11-14] with satisfying the required symmetry mentioned above. In this work, we experimentally detect such band dispersions with a characteristic spin-momentum locking, by focusing on elemental tellurium with a chiral crystal structure of space group $P3_121$ (called the "right-handed crystal") as shown in Fig. 1a,b, or $P3_221$ (called the "left-handed crystal")[15].

The crystal structure of trigonal tellurium can be illustrated as a distorted simple cubic lattice driven by the three-dimensional Peierls instability due to the two-thirds-filled orbitals in the chiral chains[16,17] with either of left-handed or right-handed structure. The right-handed crystal consisting of atomic spiral chains along the *z*-axis is characterized by the right-handed screw axes as denoted by red arrows in Fig. 1a,b. To the best of our knowledge, the band structure of tellurium has been established before the 1970s by theoretical[20-25] and experimental[18,26,27] studies, which identified a gap size of 0.32 eV[18,19] and the valence band maxima (VBM) located slightly off the *H*- and *H'*-point (Fig. 1f). The spin degeneracy at the top of valence bands which crosses the Fermi level is lifted[18] because of the combination of the SOI of tellurium 5*p* orbitals and the noncentrosymmetric crystal structure; this property could trigger the electromagnetic effects (e.g., the current-induced optical activity[29] and magnetization in bulk[30]).

Recently, Nakayama *et al*. observed the valence band structure by ARPES, which agrees with the calculational results of the density fractional theory[28]. Furthermore, the topological nature of the band structure was theoretically investigated by Hirayama *et al*. The Weyl semimetal phase was suggested to emerge when the spin-polarized conduction and valence bands around the Brillouin zone (BZ) corners are inverted across the band-gap under the hydrostatic pressure, which shows the root of exotic phenomena induced by the spin-momentum locked electronic structure[25,31].

Here, we investigate the spin and electronic structures of elemental trigonal tellurium by using ARPES and SARPES in combination with the first-principles calculations, and reveal the relationship between the chiral crystal structure and the spin polarization around the top of the valence bands. Single crystals of tellurium (Fig. 1c) were grown by the physical vapor transport method (see method). The hole carrier densities were estimated from the Hall coefficient to be in the *p*-type semiconducting regime of $1.6 \times 10^{15}$ - $6.0 \times 10^{17}$ cm$^{-3}$, which is consistent with the previous studies[18]. The chirality of the crystal was determined by observing the asymmetric etch pits formed by the hot concentrated sulfuric acid, as represented by red frames in Fig. 1d,e for the right-handed (space group $P3_121$) and left-handed ($P3_221$) crystals, respectively[32-34]. The samples were cleaved *in situ* along the (10-10) plane. While the band structure obtained by ARPES is for the bulk state, a projected two-dimensional BZ for the cleavage plane is also denoted for convenience in Fig. 1f (a blue rectangle) together with the main three-dimensional BZ (black tick lines). The $h\nu$-dependent ARPES measurements ($h\nu$ = 63 - 90 eV) were performed along the purple cut ($\bar{\Gamma}$-$\bar{H}$) on the two-dimensional BZ (Fig. 1f), which traces the shaded (Γ-*M*-*H*) plane in the three-dimensional BZ (see Fig. 1f,g).

Figure 2a-f show the result of $h\nu$-dependent ARPES measurements along $\bar{\Gamma}$-$\bar{H}$ for the left-handed crystal. We note that the band structures of the right- and left-handed crystals should be the same except for the difference of the spin components since they are connected by the mirror operation with the time-reversal symmetry. These ARPES images clearly exhibit $h\nu$ dependence (or $k_x$ dependence) of the band dispersions, which thus indicates the three-dimensional bulk state in our data. The top of the energy dispersion at $\bar{H}$-point shifts up toward higher energies with decreasing $h\nu$, and it crosses the Fermi level at $h\nu$ = 63 eV (Fig. 2f). This indicates that the *H*-point in the three-dimensional BZ (Fig. 1f) is reached at $h\nu$ = 63 eV as shown in Fig. 1g, which is consistent with a previous study using $h\nu = 62$ eV[28]. The hole-like band near the Fermi level is further examined in Fig. 2g-i by mapping the ARPES intensities crossing the *H*-point ($h\nu = 63$ eV) at several energies ($E - E_F = 0$, −0.15, and −0.30 eV) indicated by the red broken lines in Fig. 2m. For a better understanding of these, the corresponding calculational results are also plotted next to each panel of ARPES data in Fig. 2j-l; here the career number is fixed to the experimentally obtained largest value of $6.0 \times 10^{17}$ cm$^{-3}$, which locates the Fermi level 20 meV below the VBM. At $E - E_F = 0$ (Fig. 2g), faint but finite ARPES intensities are observed in a small region around the *H*-point, which is compatible with the formation of a tiny Fermi pocket with a dumbbell shape expected by calculations (Fig. 2j). The equi-energy surface becomes larger at lower energies owing to the hole-type shape of the band. In addition, the inner intensities represented by a red dashed circle (Fig. 2h) seem to appear at $E - E_F = -0.15$ eV, and the consequent double-contours get clearer at deeper binding energy ($E - E_F = -0.30$ eV; Fig. 2i). These observations close to $E_F$ agree well with the calculational results (Fig. 2k,l).

The overall hole-like band dispersions across the *H*-point are more clearly demonstrated in Fig. 2m with the $E$-$k_z$ image. The highest valence band has the energy maximum slightly off the *H*-point, and it crosses the Fermi level, whereas the top of the second-highest band reaching about $E - E_F = -0.15$ eV is situated just at *H*-point. The hole-like dispersions observed down to $E - E_F = -2.0$ eV (Fig. 2m) agree with the corresponding first-principles calculation (Fig. 2n), and these are assigned to six lone-pair states ($H_4$, $H_5$, and four $H_6$ bands) by considering the spin degree of freedom. The highest and second-highest bands are described by double group representations ($H_4$, and $H_5$, respectively)[18,21,22], and thus the spin degeneracy of these is lifted. While the $D_3$ symmetry requires no net spin polarization at the highly symmetrical *H*-point, non-zero spin polarization should appear off the *H*-point[25,28,30].

To discuss spin polarizations for the valence bands, we have performed SARPES experiments, which enable one to determine the electron spin polarization in three dimensions by using two sets of VLEED (very low energy electron diffraction) spin polarimeters[7,8] (see Supplementary Information for details). The photon energy we used is $h\nu = 18$ eV, in which the highest and second-highest bands were clearly observed (Fig. 3e). Figure 3b-d show the SARPES spectra for the left-handed crystal, measured at the **k**-points around the *H*-point indicated as #1-#13 in Fig. 3a,e (emission angles from -9.0 deg. to 9.0 deg.); each panel corresponds to the spin *x*, *y*, and *z* components ($S_x$, $S_y$, and $S_z$ for Fig 3a, b, and c, respectively) defined as parallel to the orthogonal axes in Fig. 1a,b. Significantly, the spectra for $S_z$ (Fig 3d) show a clear difference between the spin-up (red) and spin-down (blue) intensities, and the relationship between the two is reversed across the *H*-point (#7, or the zero emission angle). These contrasts to the other components ($S_x$ and $S_y$), which show nearly equivalent intensities for the spin-up and spin-down. Since the measured **k**-points of #1-#13 (blue circles in Fig. 3a) are aligned nearly parallel to the $k_z$ direction, our SARPES results are indicative of the formation of hedgehog-like radial spin texture around the *H*-point in the highest valence band, which is characteristic in a chiral crystal structure.

The spin-polarized band dispersion experimentally determined is exhibited in Fig. 3f, which maps the $z$ component ($P_z$) of spin polarization corresponding to the ARPES image in Fig. 3f. The red and blue circles in the panel plot the peak positions of the SARPES spectra for $S_z$ marked in Fig. 3d. The data clearly illustrates the outward and inward radial spin textures realized in the highest and second-highest bands, respectively, agreeing with the calculation result along *K-H-K* (or $k_z$ direction) in Fig. 3g. The spin polarizations are represented by the different areas of open red/blue circles in Fig. 3g, which estimate about $\pm55\%$ and $\pm90\%$ at the VBM ($\pm0.02$ Å$^{-1}$ off the *H*-point) and at the Fermi level (20 meV below the VBM), respectively.

The *H'*-point is another BZ corner inequivalent with the *H*-point, and thus the same outward radial spin textures are also expected around it for the left-handed crystal, since these two points are mutually linked by the time-reversal symmetry and $D_3$ symmetry. Our experimental results for the same left-handed crystal indeed confirmed it as demonstrated in Fig. 3h: the upper and middle SARPES spectra obtained at #5' close to the *H'*-point and #5' close to the *H'*-point, respectively, are almost identical. To demonstrate further the specific feature in chiral crystals, we have also measured a crystal with the opposite chirality (that is, a right-handed crystal). Agreeing with theory, we have revealed that the *z*-oriented spin polarizations are completely inverted from that in the left-handed crystal (see the lower spectra in Fig. 3h measured at #5), which indicates that the radial spin textures around the *H*- and *H'*-points are inward in the case of the right-handed crystal. Most importantly, our results experimentally demonstrate that the radial spin texture is indeed generated by the chirality of the crystal.

Figure 4a,b show the calculated Fermi surfaces along the $k_y$-$k_z$ plane around the *H*-point for the right- and left-handed crystals, respectively. The arrows represent the directions of the spin polarization projected on the *y-z* plane along the Fermi surface. The upper and lower numbers nearby each arrow describes the absolute values of the spin polarizations projected on the *y-z* plane ($|P_{yz}|$) and along the *x* direction ($P_x$), respectively. In both of the right- and left-handed crystals, the spin textures are the same between the momentum regions around the *H*- and *H'*-points, which are connected by the time-reversal symmetry. The radial spin textures are inward and outward in the right- and left-handed crystals, respectively. Depending on the crystal chirality, the BZ corners (the *H*- and *H'*-points), therefore, behave as if being either a sink or source of the effective magnetic field like a monopole.

As mentioned, the spin polarizations of about $\pm$ 90% nearly parallel to the *z* direction are realized along the longer direction of the dumbbell-shaped Fermi surface. In contrast, the spin polarizations along the shorter direction are non-zero but small. One can physically say that the structure of the tellurium atomic chain is like a solenoid, and an electron travelling along the $k_z$ direction feels an effective magnetic field along the *z*-direction. Thus, the non-zero spin components parallel to the momentum all-around the highly symmetrical points in **k**-space must be due to the peculiar symmetry of the chiral crystal structure lacking the mirror symmetry. In addition, it is also an intriguing feature of the chiral crystal that the inward and outward radial spin textures in Fig. 4a,b cannot be connected to each other by the product of rotation operations but by the inversion or mirror operation.

Finally, we discuss the spin and electronic structures just at *H*- and *H'*-point. With the $D_3$ symmetry, the net spin polarization should be absent in **k**-space. However, the SOI still plays an essential role in coupling the spin and orbital degrees of freedom. Actually, the degeneracy of the spin is lifted into two irreducible representations of the double groups $H_4$ and $H_5$ in Fig. 4c. Figure 4d,e show the real-space views of the spin texture on the atomic spiral chain of tellurium symmetrically allowed under the $D_3$ symmetry for the highest and second-highest valence bands at the *H*-point for the right-handed crystal. The in-plane outward (inward) radial spin texture is realized vertically to the *z*-axis for $H_4$ ($H_5$) state, and it behaves as if consisting of local spin magnetic moments denoted by the red and blue arrows in Fig. 4d (Fig. 4e), while the net spin polarization is canceled to be zero. On the other hand, the spin textures for electronic structure just at the inequivalent *H'*-point should be inverted from those around the *H*-point due to the time-reversal symmetry, and again, no net magnetic moment is obtained in real- and **k**-spaces. Such lifting of the spin degeneracy without net spin polarization universally occurs via SOI in a material belonging to the $D_3$ point group.

To summarize, the spin textures of the elemental trigonal tellurium with left- and right-handed chiralities were uncovered by means of ARPES and SARPES, in combination with the first-principles calculations. The observed inward or outward radial spin textures around the VBM along the spiral chains of tellurium atoms showed a one-to-one correspondence to the chirality of the crystal structure. Because the number of the spin-polarized electrons at the Fermi level is easily controlled by the hole-doping, the elemental trigonal tellurium is an ideal material with the functionality to take advantage of the interplay between the spin and charge degrees of freedom.

## Methods
### Angle-resolved photoemission spectroscopy

$h\nu$-dependent ARPES measurement ($h\nu$ = 63-93 eV) was performed by using the VG-Scienta SES2002 electron analyzer at BL28 in Photon Factory, KEK, with a total resolution energy of 25-40 meV. Samples were cleaved *in situ* and measured at 20 K.

### Crystal growth

Single crystals of tellurium (Te) were grown by the physical vapor transport (PVT) technique[35]. Te grains (6N purity) of ~0.5 g sealed in an evacuated quartz tube (~φ11mm × φ9mm × 120 mm) was heat-treated in a tube furnace with two heating zones. Temperatures were set to 450 $^{o}$C and 360 $^{o}$C at the source (Te grains) and the growth zones, respectively. As shown in Fig. 1c, after ~70 hours of the PVT, ~3 mm long Te single crystals having a hexagonal prism shape similar to a quartz crystal were obtained.

### First principles calculation

We calculate the band structures within the density functional theory in the local density approximation. We calculate the fully-relativistic electronic structure by a first-principles code QMAS (Quantum MAterials Simulator) based on the projector augmented-wave method. The plane-wave energy cutoff is set to 40 Ry, and the 6×6×6 k-mesh is employed. We evaluate the self-energy correction in the GW approximation (GWA) [36] using the full-potential linear muffin-tin orbital code[37, 38]. In the GWA, we neglect the spin orbit interaction. The 6×6×6 k-mesh is sampled and 51×2 unoccupied conduction bands are included, where ×2 is the spin degrees of freedom. We construct 9×2 maximally localized Wannier functions (MLWF's) originating from the 5p orbitals and diagonalize the fully-relativistic Hamiltonian with the GW self-energy correction expressed in the MLWF basis.

Acknowledgements

We thank T. Itou and T. Furukawa for fruitful discussion. This research was partly supported by CREST projects (Grant No. JPMJCR16F2, No. JP-MJCR14F1) from Japan Science and Technology Agency (JST) and JSPS KAKENHI (Grants-in-Aid for Scientific Research) (Grant No. JP18H03678, JP19H00651).

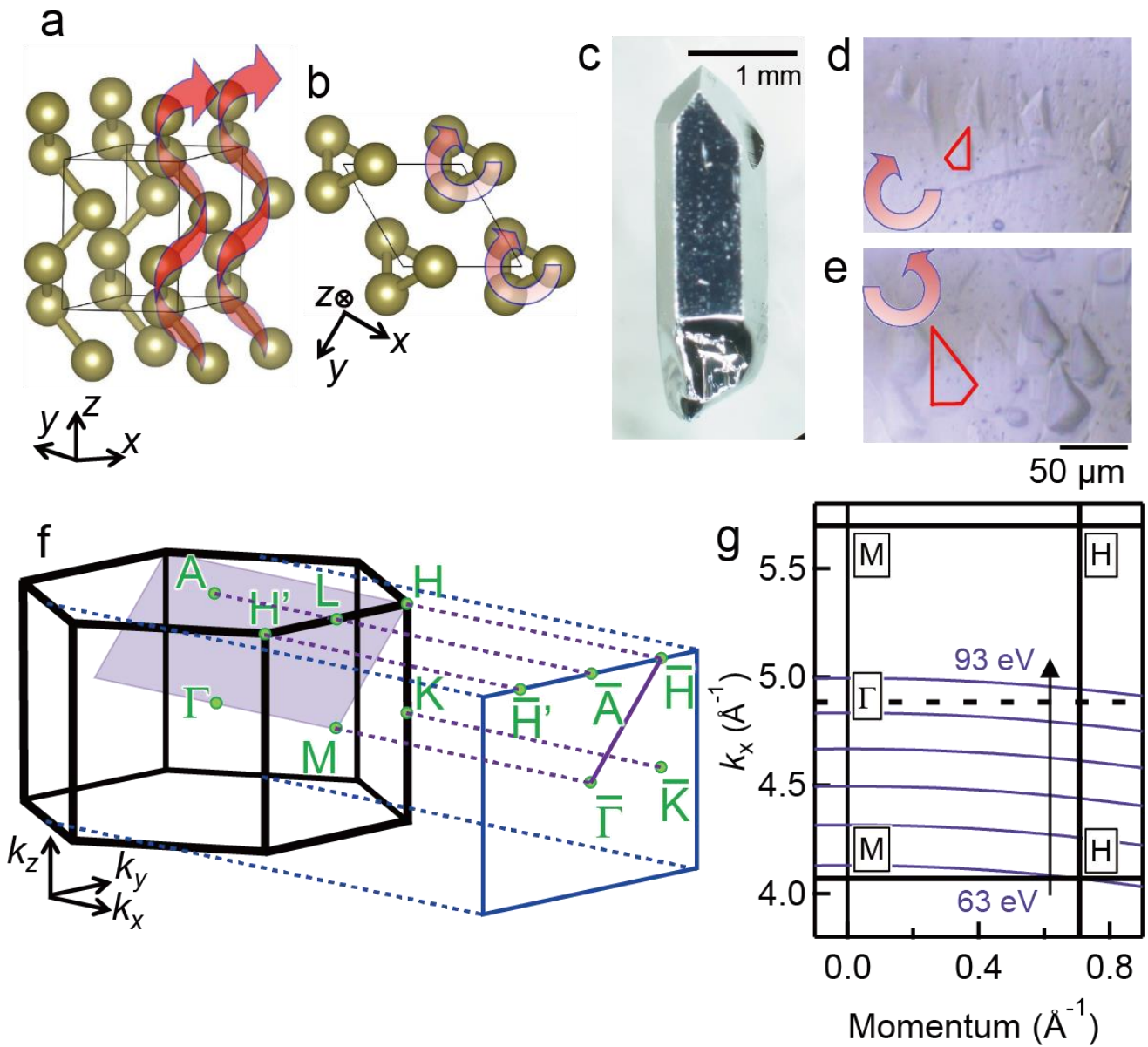

**Figure 1 | Chiral crystal structure and Brillouin zone of elemental trigonal tellurium**

**a,** Crystal structure with a right-handed screw axis (the right-handed crystal; space group $P3_121$) denoted by red arrows. **b,** Bottom view of the right-handed crystal. **c,** Picture of the right-handed single crystal. **d,e,** Optical micrograph images of etch pits represented by the red shapes for the right- and left-handed (space group $P3_221$) crystals. The right- and left-handed screw axes are denoted by the red arrows in d and e, respectively. **f,** The first Brillouin zone (thick black lines), and the corresponding two-dimensional one projected on the cleavage plane (thin blue lines). The purple line ($\bar{\Gamma}$-$\bar{H}$) indicates the orientation of measurement by angle-resolved photoemission spectroscopy (ARPES) in Fig. 2. The shaded region ($H$-$M$-$\Gamma$) can be traced by changing photon energies ($h\nu$s). **g,** Momentum cuts (purple lines) for various $h\nu$s ($h\nu$ = 63 - 93 eV) drawn on the shaded momentum plane in **f**.

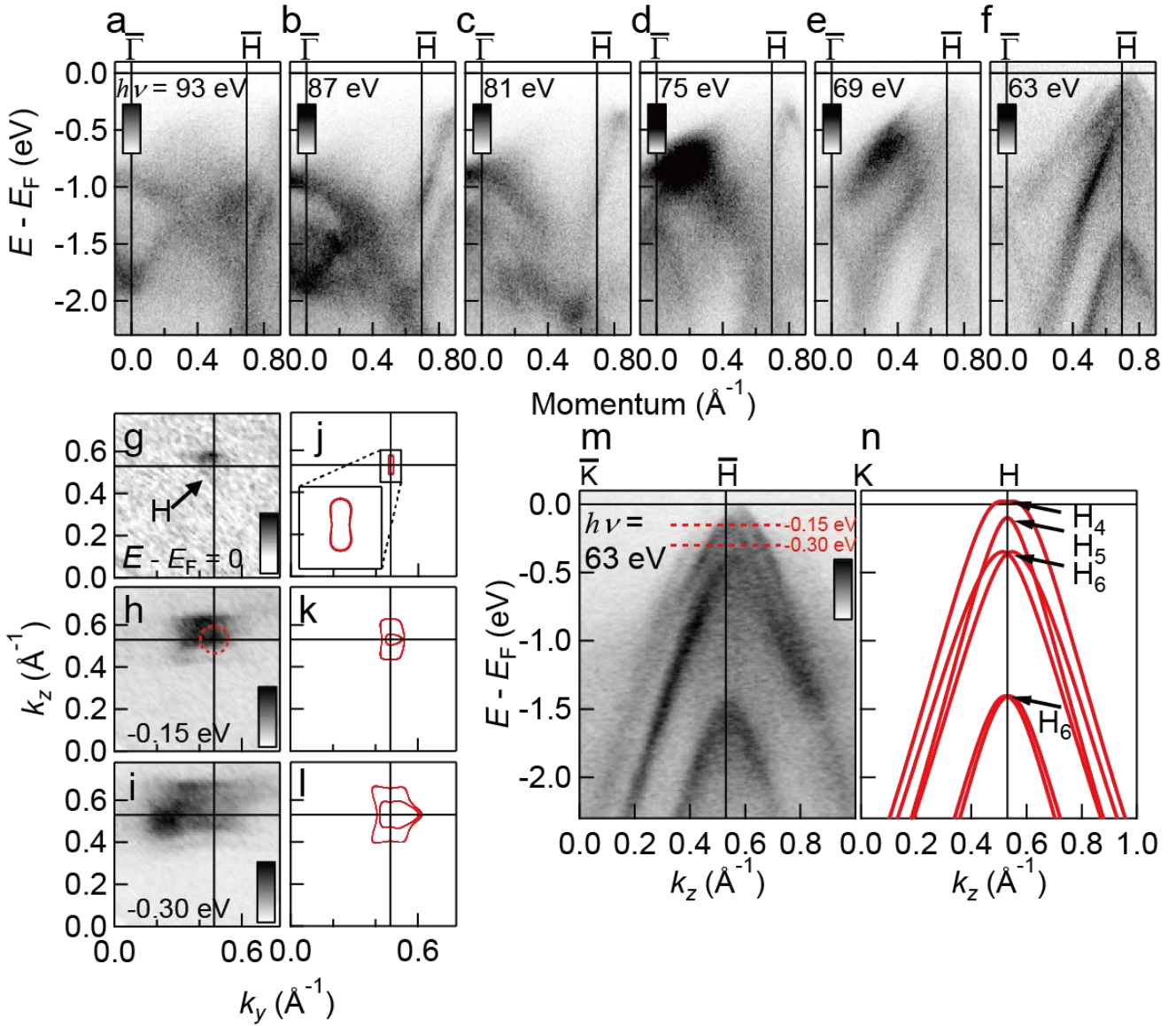

**Figure 2 | Bulk band structures of elemental trigonal tellurium**

**a-f,** The $h\nu$-dependent ARPES images along $\bar{\Gamma}$-$\bar{H}$ measured over a wide range of $h\nu$ = 93 - 63 eV at an interval of 6 eV. The corresponding momentum cuts are represented in Fig. 1g. **g-i,** ARPES intensity map along $k_y$-$k_z$ at the constant energies of $E - E_F$ = 0, -0.15, and -0.30 eV, respectively. The data are recorded at $h\nu$ = 63 eV, crossing the *H*-point (Fig. 1g). **j-l,** Energy contours of calculated valence bands corresponding to **g-i**, respectively; here the Fermi level is set to 20 meV below the VBM for the career number to be 6.0 x $10^{17}$ cm$^{-3}$. The inset in **j** shows a magnified view of the Fermi surface. **m,** $E$-$k_z$ ARPES image around the valence band maximum recorded at $h\nu$ = 63 eV. **n,** Calculated band dispersions along *K-H-K*.

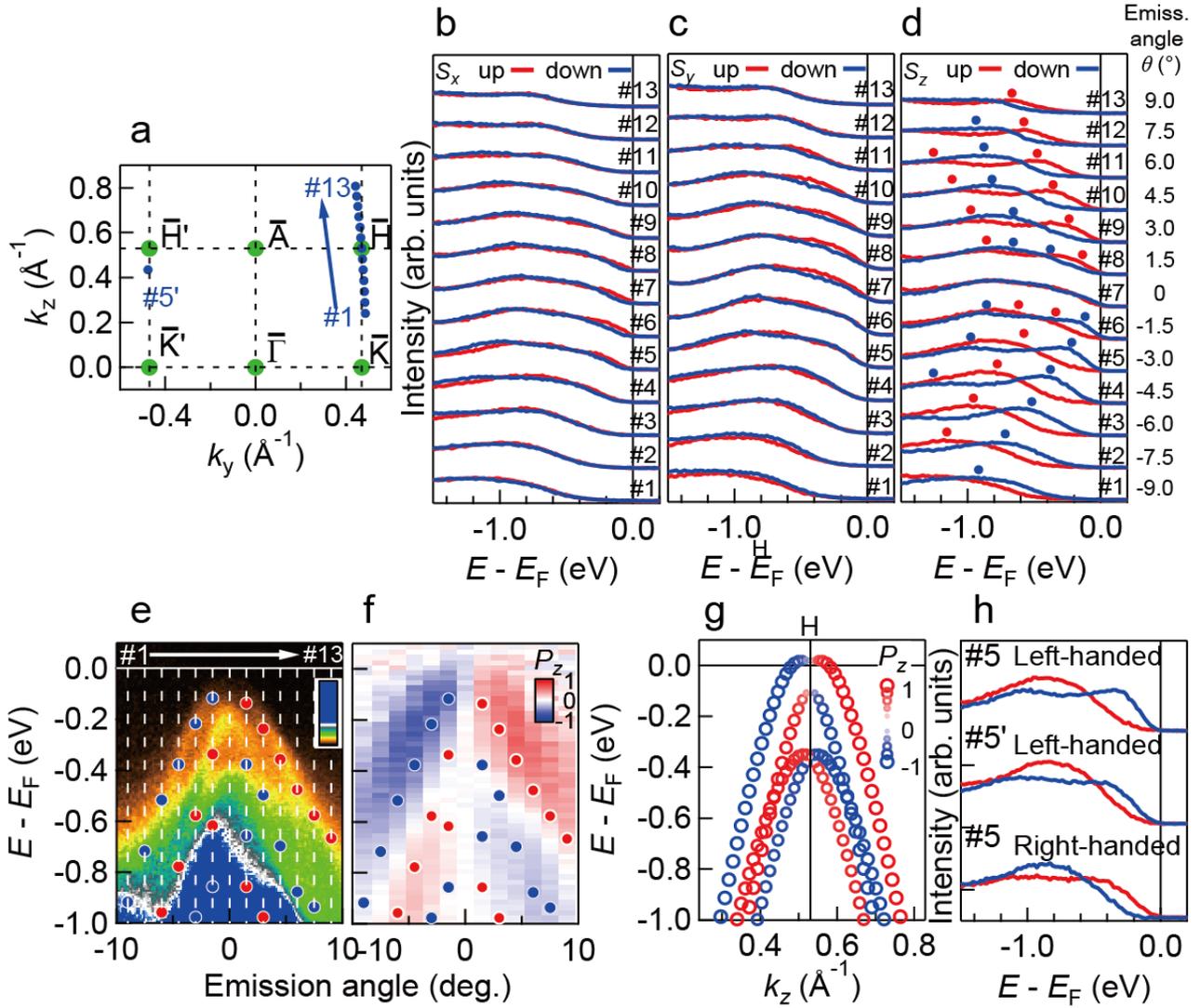

**Figure 3 | Detection of the spin polarizations near the valence band maximum around the Brillouin zone corners**

**a**, The projected two-dimensional Brillouin zone in Fig. 1f with the **k**-points (#1-#13, and #5' marked by blue circles) measured by the spin-resolved ARPES. **b-d,** Spin-resolved energy distribution curves (EDCs) for the left-handed crystal recorded at $h\nu = 18$ eV for the spin-components along the $x$, $y$, and $z$ crystal axes defined in Fig. 1a ($S_x$, $S_y$ and $S_z$), respectively. Each EDC is labeled by the measured **k**-points of #1-13 in **a**. Hereafter, red (blue) color indicates the spin-up (down) component. The markers of red and blue circles indicate the positions of the intensity peaks. **e**, ARPES image around the valence band maximum recorded at $h\nu = 18$ eV. White lines represent the measurement cuts for the EDCs in **d**. The markers indicate the peak positions of the spin-resolved EDCs, which are also plotted in **d**. **f,** Spin-resolved ARPES image for the spin $z$ component ($P_z$) of spin polarization for the left-handed crystal. The markers are the same as those in e, indicating the peak positions of the spin-resolved EDCs. **g,** Calculated spin polarization along $k_z$ ($P_z$) for the valence bands. **h,** Spin-resolved EDCs at #5 (around $H$-point) and #5' (around $H'$-point) for the left-handed crystal, and at #5 (around $H$-point) for the right-handed crystal, respectively. The **k**-points of #5 and #5' are shown in a.

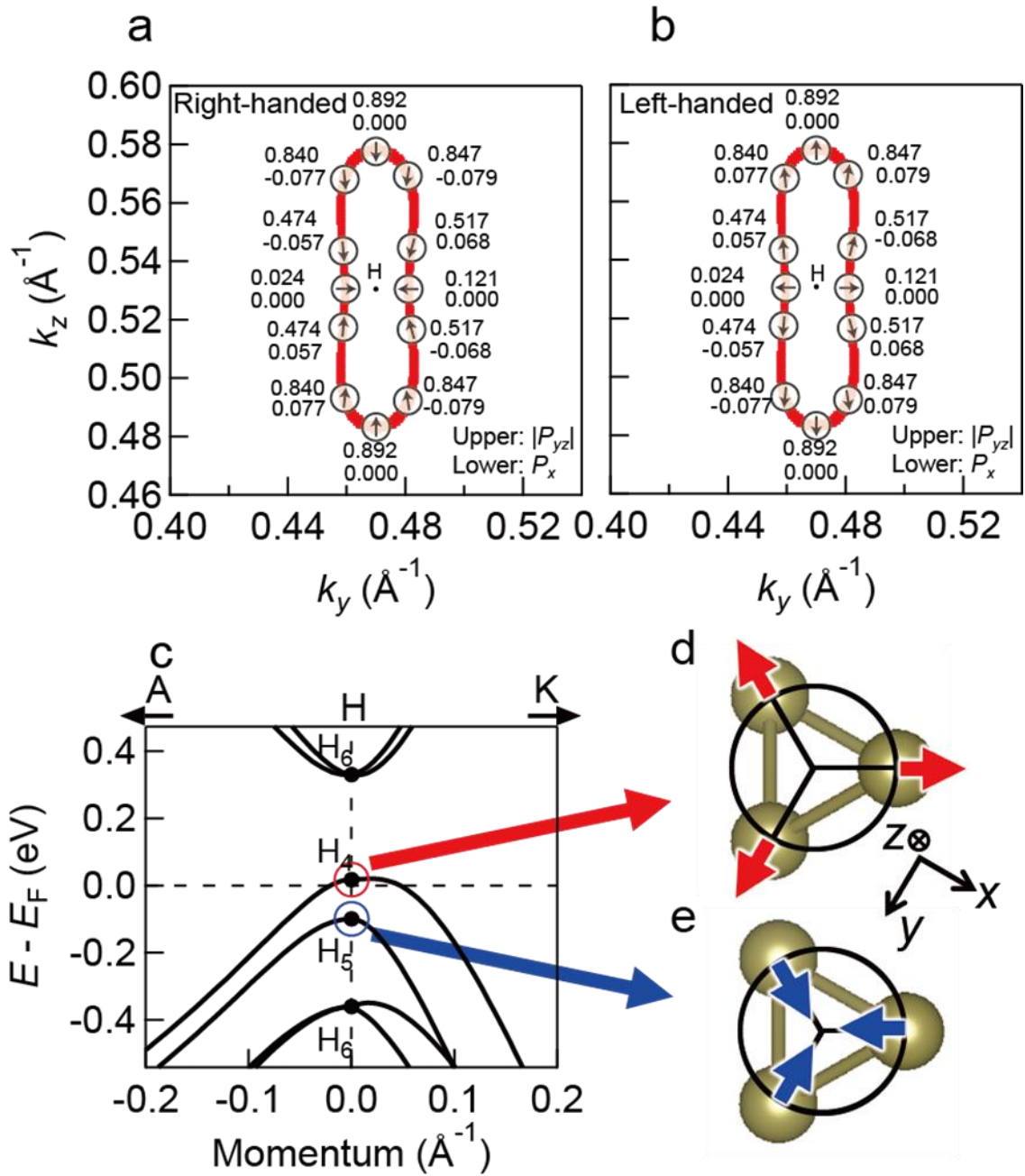

**Figure 4 | Spin textures for the right- and left-handed crystal structures in *k*- and real-spaces**

**a,b,** Calculated Fermi surfaces along $k_y$-$k_z$ formed 20 meV below the VBM around the *H*-point for the right- and left-handed crystals, respectively. The arrows mapped along the Fermi surfaces represent the directions of the in-plane spin polarization. The upper and lower numbers nearby the arrows indicate the absolute values of the in-plane spin polarization ($|P_{yz}|$) and the spin polarization along the *x*-direction ($P_x$), respectively. **c,** The calculated band dispersions around the *H*-point. **d,e,** Top views of the atomic spiral chain of tellurium. Red and blue arrows represent the symmetrically allowed spin polarizations just at the *H*-point for the eigenstates $H_4$ and $H_5$, respectively, mapped on the real space for the right-handed crystal. It is indicated that no net spin polarization is allowed at the *H*- and *H'*-points both in real- and **k**-space.

15 / 15